\documentstyle[12pt]{article}
\begin{document}

\title{Flavor Alignment Solutions to the Strong CP Problem in
Supersymmetry\footnote{Supported 
in part by Department of Energy grant
\#DE-FG02-91ER406267}}

\author{{\bf S.M. Barr}\\Bartol Research Institute\\
University of Delaware\\Newark, DE 19716}

\date{BA-97-16}
\maketitle

\begin{abstract}

An approach to solving the Strong CP Problem in supersymmetric
theories is discussed which uses abelian family symmetries
to align the mass matrices of the quarks and squarks. In this way
both the Strong CP Problem and the characteristic flavor and CP 
problems of supersymmetry can be solved in a single way.

\end{abstract}

\newpage

It is well-known that low-energy supersymmetry exacerbates the
``flavor problem".$^1$ First, there are new 
contributions to various flavor changing processes.$^2$ In particular,
since there is in general no GIM mechanism in the squark sector, 
$\Delta S = 2$ box diagrams involving squarks and gluinos can 
lead to excessive $K^0 - \overline{K^0}$ mixing. Second,
there are one-loop contributions to the u and d quark electric 
dipole moments, coming from new CP-violating phases that appear 
in the soft terms that break supersymmetry.$^3$ These contributions 
are naturally two orders of magnitude larger than the experimental bounds.
And, third, there are new contributions to $\overline{\theta}$ from diagrams
involving
gluinos and squarks. These create difficulties for non-axion solutions 
to the Strong CP Problem.$^4$
 
These various problems have led to a renewed interest in flavor symmetry$^5$
and in spontaneously-broken CP symmetry$^{6,7,8,9}$ as a way to control 
excessive violations of flavor and CP in the supersymmetrized standard model. 
Significantly, even before the advent of supersymmetry, it was suggested 
that a combination of flavor symmetry and spontaneously broken CP could 
solve the Strong CP Problem. Various models were proposed$^{10}$ that 
implemented this idea. It therefore seems reasonable in the context
of supersymmetry to attempt to find 
a unified approach to all the problems of flavor and CP violation, or 
in other words, to treat the Strong CP Problem not as a separate problem 
requiring a separate solution (the axion), but as a particularly severe 
aspect of a more general problem.$^{6,9}$ An advantage of such a combined 
approach is that it may lead to a more constrained set of possible 
solutions, and perhaps even to a unique one.

In a recent paper$^9$ we showed that an abelian flavor symmetry can
cause a ``flavor alignment"$^{11}$ of the quarks and squarks that makes
the supersymmetric contributions to $\overline{\theta}$ sufficiently
small. However, the example presented there did not deal with the
other aspects of the flavor problem. Here we propose a model, or rather
a class of models, which does, and which also has the virtue of more 
comfortably satisfying the bound on $\overline{\theta}$. 

The class of models that we propose is characterized by the following
mass matrices:

\begin{equation}
M_u = \left( \begin{array}{ccc}
s_{11} & s_{12} & s_{13} \\
0 & s_{22} & 0 \\
0 & 0 & s_{33} 
\end{array}
\right) \langle H_u \rangle
\sim \left( \begin{array}{ccc}
\lambda^6 & \lambda^4 & \lambda^3 \\
0 & \lambda^3 & 0 \\
0 & 0 & 1
\end{array} \right) v,
\end{equation}

\noindent
and 

\begin{equation}
M_d = \left( \begin{array}{ccc}
s'_{11} & 0 & 0 \\
0 & s'_{22} & s'_{23} \\
0 & 0 & s'_{33} 
\end{array}
\right) \langle H_d \rangle \sim
\left( \begin{array}{ccc}
\lambda^6 & 0 & 0 \\
0 & \lambda^4 & \lambda^4 \\
0 & 0 & \lambda^2 
\end{array} \right)v'.
\end{equation}

\noindent
Here $s_{ij} = \lambda_{ij} \langle S_{ij} \rangle/M$, and $s'_{ij} =
\lambda'_{ij} \langle S'_{ij} \rangle/M$, where $\lambda_{ij}$
and $\lambda'_{ij}$ are dimensionless effective coupling constants,
and $S_{ij}$ and $S'_{ij}$ are chiral superfields which are singlets 
under the Standard Model gauge group, and which get vacuum expectation 
values that break the flavor group. For now we will treat these 
singlets as distinct fields, in which case there are at least nine
such fields ({\it cf.} Eqs. (1) and (2)). The scale of flavor-breaking 
is $M$, which
is assumed to be somewhat below the Planck scale, but far above the
Fermi scale. $\lambda$ is the Wolfenstein parameter ($\approx 0.2$). 
These nonrenormalizable Yukawa terms come from integrating out
states of mass $M$ in a renormalizable theory.$^9$ It is also assumed that 
CP is a spontaneously broken symmetry, so that all the Yukawa couplings
are real. Both the flavor symmetry and CP invariance are 
broken by the VEVs of the singlets at the scale $M$.

If one assumes that $H_u$ and $H_d$ do not transform under the
flavor symmetry and that the $S_{ij}$ and $S'_{ij}$ are all distinct
fields, then the flavor symmetry enjoyed by these Yukawa terms is 
$\tilde{G}_F = U(1)^9$, corresponding
to rotating the phases of the nine quark fields, $Q_i$, $U^c_i$, and 
$D^c_i$ ($i= 1,2,3$), independently. Let $U(1)_{(q_i, u_i, d_i)}$ be the
particular $U(1)$ subgroup of $\tilde{G}_F$ under which these quark 
fields have charges $q_i$, $u_i$, and $d_i$ respectively. Then the 
eight $U(1)$'s that satisfy $\sum_i (2 q_i + u_i + d_i) = 0$ will
have no $SU(3)_c^2 \times U(1)$ anomaly. Let us call this 
color-anomaly-free $U(1)^8$ flavor group $G_F$. 

It is clear that there are two non-trivial combinations of the singlet
fields $S_{ij}$ that are $G_F$-invariant. (By non-trivial we mean
to exclude such combinations as $S_{ij} S_{ij}^*$.) These are

\begin{equation}
c \equiv s_{33}^{'} s_{23}^{'*} s_{22} s_{12}^{*} s_{13} s_{33}^{*} 
\sim \lambda^{16},
\end{equation}

\noindent
and

\begin{equation}
d \equiv s_{11} s_{22} s_{33} s'_{11} s'_{22} s'_{33},
\end{equation}

\noindent
where the fields have been divided by $M$ to make the quantities
$c$ and $d$ dimensionless.
If the full group $G_F$ were gauged, then the only two physically
meaningful phases in the theory would be those of $c$ and $d$.
The latter of these appears in the expression for the determinant of
the tree-level quark mass matrices. We assume that the minimization
of the Higgs potential leads $d$ to be real, at least at tree level.
Then $\overline{\theta} = 0$ at tree level. 

The CP violation in such a model comes exclusively, therefore, from the 
phase of $c$. It is because of this, and because $c$ is so high
order in $\lambda$, that it will turn out that $\overline{\theta}$
is sufficiently small. On the other hand, the Kobayashi-Maskawa
phase $\delta$ is of the same order as ${\rm arg}(c)$, which will be
assumed to be of order unity. 
This is easy to see from the fact that the invariant 
combination of KM elements $V_{td} V_{ts}^* V_{cs} V_{cd}^*$ is given
to leading order in $\lambda$ by $(s_{13}/s_{33})(s'_{23}/s'_{33})^*
(1)(s_{12}/s_{22})^*$ which is in turn equal to $c/(\left| s_{33}
s_{22} s'_{33} \right|^2)$. Thus its phase is simply the phase of $c$.

To estimate the radiatively induced value of $\overline{\theta}$ it is
necessary to examine the squark mass matrices. Assuming for the time
being the flavor group to be $G_F = U(1)^8$, 
the left-right squark mass$^2$ matrices
have the same forms as the quark mass matrices. That is, 
$(M_{LR}^{d2})_{ij} \sim A (M_d)_{ij} \sim A v' s'_{ij}$, and similarly
for the up quark sector. Of course there can be other, non-holonomic
contributions to the left-right squark masses coming from a variety
of sources.$^8$ But these will either suppressed by powers of $\langle
S_{ij}^{(')} \rangle/M_{Pl}$, or by powers of the $s_{ij}^{(')}$ and
hence high powers of $\lambda$. 

The left-left mass$^2$ matrix $M_{LL}^{d2}$ has the form 

\begin{equation}
(M_{LL}^{d2})_{ij} = a_{Li} \delta_{ij} m_0^2 + (M_d M_d^{\dagger})_{ij}
+ O(\ln (M/M_W)/16 \pi^2) (M_u M_u^{\dagger})_{ij}.
\end{equation}

\noindent
The first term represents the diagonal terms, which do not break
the flavor group $G_F$, and hence are unsuppressed. The $a_{Li}$ are
dimensionless numbers of order unity, which have no reason to show any
degeneracy. The second term is just the supersymmetric contribution.
The third term results from loops involving charged Higgs. Thus, while
the diagonal entries are of order unity times the square of the 
SUSY-breaking scale, the $(ij)$ element, where $i \neq j$, is 
proportional either to factors of $s_{ik}^{'} s_{jk}^{'*}$ or to loop factors
times $s_{ik} s_{jk}^*$, and therefore to powers of the Wolfenstein
parameter $\lambda$. The same discussion applies to the matrix
$M_{LL}^{u2}$, with the roles of $s$ and $s'$ interchanged.

The right-right mass$^2$ matrices of the squarks have analogous forms.

\begin{equation}
(M_{RR}^{d2})_{ij} = a_{Ri} \delta_{ij} m_0^2 + (M_d^{T} M_d^*)_{ij},
\end{equation}

\noindent
with a similar expression for the up quark sector. Here there are no
one-loop corrections analogous to the third term in eq. (5).

There may be contributions to the squark mass$^2$ matrices which have
a different form, especially if only a subgroup of $G_F$ is gauged,
so that other invariants than $c$ and $d$ are allowed by local symmetry.
However, if induced by Planck-scale physics, these contributions will
be suppressed by powers of $M/M_{Pl}$, which we are taking to be small.
If they are induced by loops at the scale M, they will derive from the
forms given in Eqs.(1) and (2), and thus will involve no CP-violating 
flavor-invariant except $c$. 
Therefore, in looking for the leading contribution to
$\overline{\theta}$, the forms given in Eqs. (5) and (6) are sufficient.

The leading contribution to $\overline{\theta}$ comes from the diagram
in Fig. 1(a). If one ignored flavor violation in the left-left
and right-right squark mass$^2$ matrices, the contribution of this graph to
$M_d$ would be of the form $\delta M_d = O(\alpha_s/4 \pi) 
(M_{LR}^{d2}/m_0)$$f(m_{\tilde{g}}/m_0)$. But since this has the
same form as $M_d$ itself, this gives no contribution to 
$\overline{\theta}$. Indeed, it is clear from the fact that the only 
CP-violating
invariant, $c$, involves elements of both $S_{ij}$ and $S'_{ij}$, that
one must take into acount the piece of the $M_{LL}^{d2}$ matrix that
involves $M_u$, namely the third term in Eq. (5). Effectively, then,
$\overline{\theta}$ is a two-loop effect. This is a central idea behind 
these models
and of the forms given in Eqs. (1) and (2). Because the invariant that 
violates CP involves both $M_d$ and $M_u$, the exchange of charged
states, either $W^{\pm}$ or $H^{\pm}$ is required to bring it into
play, thus necessitating higher loops.

When one includes the effect of the third term of Eq. (5) in the 
diagram of Fig. 1(a), one finds straightforwardly that

\begin{equation}
\delta \overline{\theta} \sim \left( \frac{\alpha_s}{4 \pi} \right)
O \left( \frac{\ln 
(M^2/M_W^2)}{16 \pi^2} \right) (c/\left| s'_{33} \right|^2)
\stackrel{_<}{_\sim} 3 \times 10^{-3} \lambda^{12} \stackrel{_<}{_\sim} 
10^{-10}.
\end{equation}

\noindent
The analogous contribution to $M_u$ gives a smaller contribution to
$\overline{\theta}$. 

There is also a contribution to $\overline{\theta}$ from the diagram
in Fig. 1(b). It is straightforward to see that this gives 
$\delta \overline{\theta} \stackrel{_<}{_\sim} (\alpha_s/4 \pi)
(A/m_{\tilde{g}}) (v'/m_0)^2 {\rm arg}(c) \sim 10^{-2} \; 
\lambda^{16}/\tan^2 \beta \sim 10^{-12}/\tan^2 \beta$. This is evidently
much smaller than the contribution from Fig. 1(a).

The problem of excessive flavor changing in supersymmetric models is 
here solved in the same way as in the models of ``flavor alignment" 
proposed by Nir and Seiberg.$^{11}$
In particular, the danger of excessive $K^0- \overline{K^0}$ mixing 
coming from gluino box diagrams is obviated by the absence of a 12 
element in $M_d$. This means that, as in the Nir-Seiberg models,
the Cabibbo mixing must come from the up-quark sector, which in turn
implies that the mixing in the $D^0-\overline{D^0}$ sector is near
the experimental bound.$^{11}$

Finally, there is the question of excessive electric dipole moments
(or chromo-electric dipole moments) for the $u$, $d$, or $s$ quarks.
It is easy to see that these, since they also must involve
the invariant $c$, are suppressed by large powers of $\lambda$. 
In fact they are less than or of order $e (\alpha_s/4 \pi) 
(A m_{\tilde{g}} v^{'2} v/m_0^6) \lambda^{16}$, which is less than
$10^{-28}$e-cm, or about
three orders of magnitude below the experimental bound.

The pattern or ``texture" given in Eqs. (1) and (2) is unique
in the following sense. There are several other textures that
give the right amount of KM mixing, the right hierarchy of
quark masses, have vanishing $12$ element for $M_d$ in order to
avoid excessive $K^0-\overline{K^0}$ mixing, and have 
$\overline{\theta}$ vanish at tree level and suppressed by several
powers of $\lambda$ at one-loop level. However, none of them suppress
$\overline{\theta}$ by as many powers of $\lambda$ as the forms given
in Eqs. (1) and (2). There are two forms that give $\overline{\theta}$
to be of order $(\alpha_s/4 \pi) \lambda^{10}$. ({\it Cf.} Eq. (7).)
One of these is the same as the forms in Eqs. (1) and (2) except that
the $13$ element of $M_d$ rather than of $M_u$ is non-vanishing.
The other is the same as Eqs. (1) and (2) except that $M_d$ has
vanishing $23$ element and non-vanishing $13$ element, while $M_u$
has vanishing $13$ element, and non-vanishing $23$ element. Other
forms have $\overline{\theta}$ arising at even lower order in 
$\lambda$. For example, if $M_d$ has a diagonal form, and $M_u$
has a triangular form, then $\overline{\theta}$ arises at order
$\lambda^6$ as in Ref. 9.

There are many ways to construct a Higgs superpotential that
ensures that at tree level $d$ is real and $c$ complex.
An example which is easy to analyze is the following. 
Let $W_{{\rm Higgs}} = W_0 + W_d + W_c$. $W_0$ has the form
$\sum_{ij} (S_{ij} \overline{S}_{ij} - M^2_{ij}) Y_{ij} +
\sum_{ij} (S'_{ij} \overline{S'}_{ij} - M^{'2}_{ij}) Y'_{ij}$.
Here all the $M^2_{ij}$ are taken to be real and positive,
except $M^2_{13}$ which is real and negative. This ensures that
$\langle \overline{S}_{ij} \rangle = \langle S_{ij} \rangle^*$, and 
similarly for the $S'_{ij}$, except that
$\langle \overline{S}_{13} \rangle = - \langle S_{13} \rangle^*$.

$W_d$ fixes the phase of $d$ and can be taken to have the
form $\sum_k S_{kk} S'_{kk} A_k + \sum_k \overline{S}_{kk}
\overline{S'}_{kk} \overline{A}_k + A_1A_2A_3 + \overline{A}_1
\overline{A}_2 \overline{A}_3 + \sum_k m^2_k A_k \overline{A}_k$.
Integrating out the $A_k$ and $\overline{A}_k$ gives an effective
terms of the form $S_{11}S_{22}S_{33}S'_{11}S'_{22}S'_{33} \sim d$
and $\overline{S}_{11}\overline{S}_{22}\overline{S}_{33}\overline{S'}_{11}
\overline{S'}_{2}\overline{S'}_{3} \sim \overline{d}$. Together,
the conditions $F_{S_{kk}} = 0$ and $F_{\overline{S}_{kk}} = 0$, imply that
$\langle d \rangle = \langle \overline{d} \rangle = \langle d \rangle^*$ 
and therefore that $\langle d \rangle$ is real.

$W_c$ fixes the phase of $c$ and may be taken to be of the form
$W_c = S'_{33} \overline{S'}_{23} B_{23} + S_{22} \overline{S}_{12}
B_{21} + S_{13} \overline{S}_{33} B_{13} + (S \leftrightarrow
\overline{S}, B \leftrightarrow \overline{B})$. Integrating out the
$B_{ij}$, and using the equations $F_{S_{ij}} = 0$, one finds 
$\langle c \rangle = \langle \overline{c} \rangle$, in an obvious
notation. From the relation $\langle \overline{S}_{13} \rangle = 
- \langle
S_{13} \rangle^*$, it follows that $\langle \overline{c} \rangle = 
- \langle c \rangle^*$ and therefore that $\langle c \rangle$ is pure
imaginary. This is not realistic, since ${\rm arg}(c) = {\rm arg}
(V_{td} V_{ts}^* V_{cs} V_{cd}^*) = {\rm arg} (1 - \rho - i \eta)$
in Wolfenstein parametrization, and therefore ${\rm arg}(c) \neq \pi$.
But it is easy to construct superpotentials that give other phases to 
$c$.

It is possible to gauge some subset of $G_F = U(1)^8$, the full 
flavor symmetry of the quark  
Yukawa terms. Since there is no $SU(3)_c^2
\times G_F$ anomaly, by construction, the gauge anomalies can be cancelled
by auxiliary leptons, whose presence has no effect on $\overline{\theta}$.
There are nine fields, $S_{ij}$ and $S'_{ij}$, whose VEVs break
$G_F$, but two combinations of these fields, $c$ and $d$ are
$G_F$ invariant. Thus the VEVs of the singlet fields break $U(1)^8$ 
down to a
single $U(1)$ factor, which is obviously the $U(1)$ of baryon number
as far as its action on the quarks is concerned. If the broken $U(1)^7$
is gauged, and the unbroken $U(1)$ is global, there are no goldstone
bosons or pseudo-goldstone bosons associated with flavor breaking, and
all the flavor gauge bosons will have mass of order $M$, which is
safely heavy.

While the group $G_F$ is convenient for analysis, it is not
necessary that the local flavor group actually be this large. 
Nor is it necessary that there be as many singlet fields $S$
as has been assumed to this point. This is shown by the following
example which has a single gauged U(1) flavor group and six 
flavor-breaking singlet fields, but essentially the same flavor 
structure as in eqs. (1) and (2). Let there be the following singlet
fields: $S_2$, $S_3$, $S'_3$, $S_4$, $S'_4$, and $S_6$. The subscripts
correspond to the order in $\lambda$ of each field's vacuum expectation
value. For example, $\langle S'_3 \rangle /M \sim \lambda^3$. 
The quark mass matrices
have the following form (where the Yukawa couplings, assumed to be of
order unity, are not indicated)

\begin{equation}
M_u = \left( \begin{array}{ccc}
s_6 & s'_4 & s'_3 \\
0 & s_3 & 0 \\
0 & 0 & 1
\end{array} \right) \langle H_u \rangle,
\end{equation}

\noindent
and 

\begin{equation}
M_d = \left( \begin{array}{ccc}
s_6 & 0 & 0 \\
0 & s_4 & s'_4 \\
0 & 0 & s_2
\end{array} \right) \langle H_d \rangle.
\end{equation}

The Higgs superpotential can be arranged so that (in some phase
convention) the vacuum expectation value of $S'_3$ is pure imaginary
while those of the other singlets are real (as was the case in 
the previous model). These forms of the quark Yukawa
matrices can be enforced by a family $U(1)$ under which
the doublet Higgs fields, $H_u$ and $H_d$, are neutral and
the singlet fields have the charges $Q(S_2, S_3, S'_3, S_4, S'_4, S_6) = 
(x, y, z, t, \frac{1}{2}(x + y + z), -\frac{1}{2}(x + y + t))$, and 
the quark fields have the charges $Q(Q_1, Q_2, Q_3, U^c_1, U^c_2,
U^c_3, D^c_1, D^c_2, D^c_3) = (-z, \frac{1}{2}(x-y-z), 0, \frac{1}{2}
(x+y+t) +z, \frac{1}{2}(-x-y+z), 0, \frac{1}{2}(x+y+t)+ z, 
\frac{1}{2}(-x+y+z) -t, -x)$. This $U(1)$ has no $SU(3)_c^2\times U(1)$
anomaly. 

The values of $x$, $y$, $z$, and $t$ must satisfy several
conditions. In particular, the resulting charges of the fields must
be such that there are no additional terms allowed in the matrices
in Eqs. (8) and (9). The zeros must stay zeros, and the non-zero entries 
must arise from a single field. Moreover, the Higgs superpotential
must contain enough distinct kinds of terms to prevent unwanted
accidental global flavor symmetries, but no terms which make the
invariant $d$ have a complex vacuum expectation value. There are many
solutions to these conditions. One example is $(x,y,z,t) =
(+1,-1,-4,-6)$. This allows the terms $S_2 S_3$, $S_4 S_6 S_6$,
$S_2 S_2 S'_4$, and $S'_3 S_2 S_6$ to appear in the superpotential, 
which thus prevents accidental flavor $U(1)$ symmetries from
arising. It is possible to construct a superpotential so that
the vacuum expectation value of each of these four invariants is
real (in which case $d$ is also, since it is the product of the
first two of them), while $c$ has a complex VEV.

There are presumably a variety of other
sets of singlets and abelian family symmetries which implement
the general Yukawa pattern of Eqs. (1) and (2). One unsatisfactory
feature of the examples presented above is that they do not 
explain the hierarchy in quark masses, as is done, for example, in
models of the Froggatt-Nielsen type,$^{12}$ where terms of higher order 
in $\lambda$ arise from higher powers of a flavor-breaking field.
There is clearly something quite {\it ad hoc} about the second
example presented. Of more significance are the general features
of these models, which it is useful to contrast with other types
of models invented to solve the strong CP problem.

One class of models, proposed almost twenty years ago$^{10}$ in a 
non-SUSY context, was similar to the kind of model proposed here 
in that they used symmetries to restrict the form of the quark mass matrices
in such a way that they had (at tree level at least) real determinants in
spite of having some elements with phases of order unity.
However, most of those models had non-minimal Higgs, and in
particular several Higgs doublets that contributed to the masses
of quarks of a given charge. This, as is well known, leads to
problems with Higgs-mediated flavor violation.$^{13}$
The same feature also typically gave rise to one-loop contributions
to $\overline{\theta}$ that tended to be somewhat too large.
With minimal Higgs structure, there are only the two Yukawa matrices,
proportional to $M_u$ and $M_d$. Thus a one-Higgs-loop contribution
to the down quark mass matrix would have the form $M_i M_i^{\dagger} M_d$,
where $i = u$ or $d$.
But then $\overline{\theta} = {\rm arg} \det (M_d + {\rm const.} M_i 
M_i^{\dagger} M_d) =  {\rm arg} [\det ({\rm Hermitian}) \cdot \det 
M_d] = 0$. The same is true for one-loop corrections to $M_u$.
But with several Higgs doublets contributing to $M_d$, as in the models
of Ref. 10, there are several
Yukawa coupling matrices, $Y^k_d$, for the down quarks.
Thus the tree plus one-loop contributions to $M_d$ have the
form $(M_d + {\rm const.} Y^k_d Y^{l \dagger}_d Y^m_d)$, which has no
reason to have a real determinant. 

An advantage of the present models consists in the fact that there
is a minimal {\it doublet} Higgs structure. Instead of there being
several Higgs doublets which couple differently in flavor and which violate
CP spontaneously, there are in
the present models several singlet scalars, $S_{ij}$ and $S'_{ij}$, which
perform the same tasks. In this respect
the models proposed here are similar to the models proposed by Nelson in 
Ref. (14). Of course, as in the models of Ref. (14), there can be one-loop
contributions to $\overline{\theta}$ coming from the emission and
reabsorption of the heavy singlet fields.
In non-supersymmetric Nelson models for such loops to be made sufficiently 
small requires
certain Yukawa couplings to be less than about $10^{-2}$ (which
is not unreasonable). Here, such loops are suppressed by
$m_{SUSY}/M$. 

The Nelson-type models have a problem, however, is the context of
supersymmetry (unless supersymmetry breaking happens at low scales and 
is
mediated by gauge interactions$^{4,9}$). The problem is that even with
minimal Higgs structure, other matrices in flavor space exist
besides the Yukawa matrices, namely the squark mass$^2$ matrices.
These allow one-loop contributions to $\overline{\theta}$ from
diagrams involving squarks and gluinos. ({\it Cf.} Fig. 1.)
In the present models these are suppressed by ``flavor alignment",
somewhat in the spirit of the old non-supersymmetric models rather
than the Nelson models. The models proposed here can therefore
be regarded as somewhat of a hybrid between the two approaches,
using features of each to suppress all one-loop contributions
to the QCD angle. 

An important feature of the ``flavor alignment" here is that
the non-zero elements in the quark mass ``textures" have a pure
form. That is, each element is generated by the VEV of a single
$S$ field. This is in contrast to both the supersymmetric Nelson
models discussed in Ref. (6) and to the models of Nir and Rattazzi.$^8$

If the flavor alignment idea in the form presented here, where all
CP violating effects come from a single flavor-invariant, $c$, of 
high order in the Wolfenstein parameter, is the true solution to the
Strong CP Problem, one would expect the following signatures:
$\overline{\theta}$ should be observed not far below the $10^{-10}$
level (compared to the value $10^{-15}$ typical of most invisible
axion models); $D^0-\overline{D^0}$ mixing should be seen not far below
the present limits; the CP violation observed in the $B$ systems should
be consistent with it all coming from the Kobayashi-Maskawa phase, 
$\delta$; the electric dipole moment of the electron should be less
than about $10^{-28}$ e-cm, and that of the neutron should come
predominantly from $\overline{\theta}$ and therefore be not much below
$10^{-26}$ e-cm.

\section*{References}

\begin{enumerate}
\item S. Dimopoulos and H. Georgi, {\it Nucl. Phys.} {\bf B193}, 150
(1981); L.J. Hall, V.A. Kostelecky, and S. Raby, {\it Nucl. Phys.}
{\bf B267}, 415 (1986); H. Georgi, {\it Phys. Lett.} {\bf 169B},
231 (1986); S. Dimopoulos and D. Sutter, {\it Nucl. Phys.} {\bf B452},
496 (1995).
\item J. Donoghue, H. Nilles, and D. Wyler, {\it Phys. Lett.} {\bf 128B},
55 (1983); F. Gabbiani and A. Masiero, {\it Nucl. Phys.} {\bf B322},
235 (1989); J. Hagelin, S. Kelley, and T. Tanaka, {\it Nucl. Phys.}
{\bf B415}, 293 (1994). For a recent comprehensive study see 
F. Gabbiani, E. Gabrielli, A. Masiero, and L. Silvestrini, hep-ph 9604387.
\item J. Ellis, S. Ferrara, and D.V. Nanopoulos, {\it Phys. Lett.}
{\bf 114B}, 231 (1982); J. Polchinski and M.B. Wise, {\it Phys. Lett.}
{\bf 125B}, 393 (1983); W. B\"{u}chmuller and D. Wyler, {\it Phys. Lett.}
{\bf 121B}, 321 (1982); E. Franco and M. Mangano, {\it Phys. Lett.}
{\bf 135B}, 445 (1984); F. del Aguila, M.B. Gavela, J.A. Grifols, and
A. Mendez, {\it Phys. Lett} {\bf 126B}, 71 (1983).
\item M. Dine, R. Leigh, and A. Kagan, {\it Phys. Rev.} {\bf D48},
2214 (1993).
\item P. Pouliot and N. Seiberg, {\it Phys. Lett.} {\bf 318B}, 169 (1993);
M. Dine, R. Leigh, and A. Kagan, {\it Phys. Rev.} {\bf D48},
4269 (1993). D.B. Kaplan and M. Schmaltz, {\it Phys. Rev.} {\bf D49},
3741 (1994); A. Pomarol and D. Tommasini, {\it Nucl. Phys.}
{\bf B466}, 3 (1996); R. Barbieri, G. Dvali,
and L.J. Hall, {\it Phys. Lett.} {\bf B377}, 76 (1996); 
L.J. Hall and H. Murayama, {\it Phys. Rev. Lett.}
{\bf 75}, 3985 (1995); C. Carone, L.J. Hall, and H. Murayama, {\it 
Phys. Rev.} {\bf D54}, 2328 (1996);
P. Frampton and O. Kong, {\it Phys. Rev.} {\bf D53}, 2293 (1996);
K.C. Chou and Y.L. Wu, {\it Phys. Rev.} {\bf D53}, 3492 (1996).
\item S.M. Barr and G. Segre, {\it Phys. Rev.} {\bf D48}, 302 (1993).
\item K.S. Babu and S.M. Barr, {\it Phys. Lett.} {\bf 387B}, 87 (1996).
\item Y. Nir and Rattazzi, {\it Phys. Lett.} {\bf 382B}, 363 (1996).
\item S.M. Barr, BA-96-56, hep-ph/9612396, to appear in {\it Phys. Rev.}
{\bf D}.
\item M.A.B. Beg and H.S. Tsao, {\it Phys. Rev. Lett.} {\bf 41}, 278
(1978); H. Georgi, {\it Had. J.} {\bf 1}, 155 (1978); R.N. Mohapatra
and G. Senjanovi\'{c}, {\it Phys. Lett.} {\bf 126B}, 283 (1978);
G. Segre and H.A. Weldon, {\it Phys. Rev. Lett.} {\bf 42}, 1191
(1979); S.M. Barr and P. Langacker, {\it Phys. Rev. Lett.} {\bf 42}, 1654
(1979).
\item Y. Nir and N. Seiberg, {\it Phys. Lett.} {\bf 309B}, 337 (1993).
\item C.D. Froggatt and H.B. Nielsen, {\it Nucl. Phys.} {\bf B147}, 277
(1979).
\item S.L. Glashow and S. Weinberg, {\it Phys. Rev.} {\bf D15}, 1958 (1977).
\item A. Nelson, {\it Phys. Lett.} {\bf 136B}, 387 (1983); S.M. Barr,
{\it Phys. Rev.} {\bf D30}, 1805 (1984).
\end{enumerate}

\newpage 

\noindent
{\bf\large Figure Captions}

\vspace{1cm}

\noindent
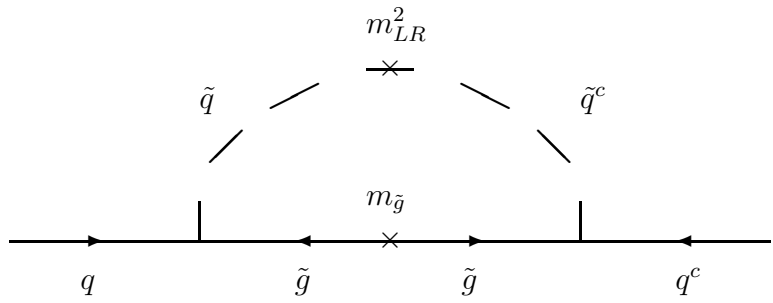
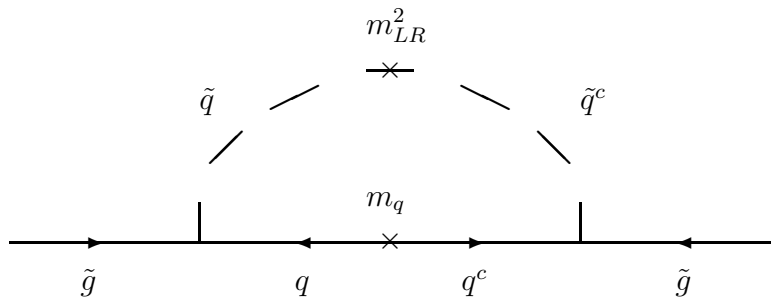
{\bf Fig. 1:} In supersymmetric models these diagrams give 
contributions to $\overline{\theta}$ through (a) the phases of
quark masses, and (b) through the phase of the gluino mass.

\newpage

\begin{picture}(360,216)
\thicklines
\put(108,72){\line(0,1){15}}
\put(112,102){\line(1,1){12}}
\put(135,122.5){\line(2,1){18}}
\put(171,137){\line(1,0){18}}
\put(207,131.5){\line(2,-1){18}}
\put(236,114){\line(1,-1){12}}
\put(252,87){\line(0,-1){15}}
\put(36,72){\vector(1,0){36}}
\put(72,72){\line(1,0){72}}
\put(180,72){\vector(-1,0){36}}
\put(180,72){\vector(1,0){36}}
\put(216,72){\line(1,0){72}}
\put(324,72){\vector(-1,0){36}}
\put(175.5,69.5){$\times$}
\put(175.5,134.5){$\times$}
\put(63,54){$q$}
\put(288,54){$q^c$}
\put(144,54){$\tilde{g}$}
\put(207,54){$\tilde{g}$}
\put(171,86){$m_{\tilde{g}}$}
\put(171,151){$m^2_{LR}$}
\put(108,122.5){$\tilde{q}$}
\put(252,122.5){$\tilde{q}^c$}
\put(160,0){\bf Fig. 1 (a).}
\end{picture}

\begin{picture}(360,216)
\thicklines
\put(108,72){\line(0,1){15}}
\put(112,102){\line(1,1){12}}
\put(135,122.5){\line(2,1){18}}
\put(171,137){\line(1,0){18}}
\put(207,131.5){\line(2,-1){18}}
\put(236,114){\line(1,-1){12}}
\put(252,87){\line(0,-1){15}}
\put(36,72){\vector(1,0){36}}
\put(72,72){\line(1,0){72}}
\put(180,72){\vector(-1,0){36}}
\put(180,72){\vector(1,0){36}}
\put(216,72){\line(1,0){72}}
\put(324,72){\vector(-1,0){36}}
\put(175.5,69.5){$\times$}
\put(175.5,134.5){$\times$}
\put(63,54){$\tilde{g}$}
\put(288,54){$\tilde{g}$}
\put(144,54){$q$}
\put(207,54){$q^c$}
\put(171,86){$m_q$}
\put(171,151){$m^2_{LR}$}
\put(108,122.5){$\tilde{q}$}
\put(252,122.5){$\tilde{q}^c$}
\put(160,0){\bf Fig. 1 (b).}
\end{picture}

\end{document}